\documentclass[12pt]{article}
\usepackage{graphicx}
\usepackage{subfigure}
\newcommand{\beq}{\begin{equation}}
\newcommand{\eeq}{\end{equation}}
\newcommand{\bea}{\begin{eqnarray}}
\newcommand{\eea}{\end{eqnarray}}

\newcommand{\ra}{\right\rangle}
\newcommand{\la}{\left\langle}

\begin{document}
\begin{titlepage}
\begin{flushleft}
       \hfill                      {\tt hep-th/0609152}\\
       \hfill                       FIT HE - 06-02 \\
       \hfill                       KYUSHU-HET ** \\
       \hfill                       Kagoshima HE - 06-2 \\
\end{flushleft}
\vspace*{3mm}
\begin{center}
{\bf\LARGE Gauge theory in de Sitter space-time \\
from a holographic model}

\vspace*{5mm}
\vspace*{12mm}
{\large Kazuo Ghoroku\footnote[2]{\tt gouroku@dontaku.fit.ac.jp},
Masafumi Ishihara\footnote[3]{\tt masafumi@higgs.phys.kyushu-u.ac.jp},
Akihiro Nakamura\footnote[4]{\tt nakamura@sci.kagoshima-u.ac.jp}
}\\
\vspace*{2mm}

\vspace*{2mm}

\vspace*{4mm}
{\large ${}^{\dagger}$Fukuoka Institute of Technology, Wajiro, 
Higashi-ku}\\
{\large Fukuoka 811-0295, Japan\\}
\vspace*{4mm}
{\large ${}^{\ddagger}$Department of Physics, Kyushu University, Hakozaki,
Higashi-ku}\\
{\large Fukuoka 812-8581, Japan\\
\vspace*{4mm}
{\large ${}^{\S}$Department of Physics, Kagoshima University, Korimoto 1-21-35,Kagoshima 890-0065, Japan\\}}

\vspace*{10mm}
\end{center}

\begin{abstract}
Yang-Mills theory with flavor quarks in the dS${}_4$
is studied through the dual supergravity in the
AdS${}_5\times S^5$ background with non-trivial dilaton and axion. The flavor
quarks are introduced by embedding a probe D7 brane. 
We find that the dynamical properties of YM theory in the dS${}_4$
are similar to the case of the finite temperature theory
given by the 5d AdS-Schwarzschild background. 
In the case of dS${}_4$, however, contrary to the finite temperature case, 
the gauge field condensate plays an important role on the dynamical
properties of quarks. 
We also give the quark-antiquark potential and meson spectra
to find possible quark-bound states. And we arrive at the conclusion that,
while the quarks are not confined in the dS${}_4$, we could find stable
meson states at very small cosmological constant as expected in the present
universe. But there would be no hadrons at early universe as in the
inflation era.

\end{abstract}
\end{titlepage}

\section{Introduction}

Recently, there have been many approaches to QCD from the gauge/gravity
correspondence based on the superstring theory \cite{MGW}. 
Especially, the idea, proposed by Karch and Katz~\cite{KK}, to add light 
flavor quarks by embedding D7 brane(s) as probe has stimulated many authors
who have developed and extended this idea to various 10d gravity solutions
corresponding to the various gauge theories. And many interesting
results have been obtained for the properties of quarks and and their 
bound states, mesons, 
in the context of the holography
~\cite{KMMW,KMMW2,Bab,ES,SS,NPR,GY,CNP}.

This approach has been developed in many directions, but the analyses are
restricted to the gauge theory in 4d Minkowski space-time and to the
finite temperature theory. On the other hand, some 
holographic approaches to the gauge theory in the 4d de-Sitter space
(dS${}_4$) are seen
\cite{Hawking,Alishahiha1,Alishahiha2,H}, but it is not enough
and we will need more study.
This situation of the positive cosmological constant
can be set for the early inflation universe or for the present
small acceleration which has been observed recently in our universe.
From this cosmological viewpoint, it would be 
important to make clear the non-perturbative properties of the 
gauge theories at finite cosmological constant.
It would be a difficult problem to see the non-perturbative behaviors
of field theories in a curved space from the standard field theory, 
so this problem is a challenging one.

Here we study this problem from the holographic approach which has been
useful 
in the finite temperature case \cite{GY}. The bulk solution corresponding 
to the gauge theory in dS${}_4$
is obtained from type IIB string theory with dilaton and axion under the
five form flux. And the  D7 brane is embedded in this background
as a probe to introduce the flavor quarks. In the bulk background, there
appears a horizon as in the finite temperature case. And we find an attractive
force between D3 and D7 branes, then the chiral symmetry is preserved
as in the high temperature phase. But a phase transition observed in the
high temperature case is seen only for the case of large gauge field 
condensate. For small gauge condensate, it is deformed to a different
situation which has not seen in the finite temperature case.

\vspace{.3cm}
In order to study the quark confinement, we obtained the potential between
quark and anti-quark through the estimation of the
Wilson-Polyakov loop. We find a maximum
distance between quark and antiquark to maintain the U-shaped string state.
Above this length, the quark and antiquark are separated as independent
particles. And the energy of each single quark (or anti-quark) state is
finite. 
This point is assured through
the estimation of the effective
quark mass. Then, in this sense, we can say that the theory is in the 
deconfinement phase. We compare these results
with the one given at the finite temperature.

Furthermore, we estimate the meson masses through the fluctuation of the
D7 brane, and we find interesting spectra for light mesons. However, all
the states would disappear at large cosmological constant since they becomes
unstable and decay to free quarks and ant-quarks.
A similar phenomenon for the baryon spectra is also
seen by studying the energy of
the D5 baryon wrapped on $S^5$ which is regarded as baryon. 

\vspace{.2cm}
In section 2, we give the setting of our model for our study, 
and a problem of the phase transition is discussed by 
solving the embedding of the D7 brane. 
In section 3, effective mass of the quark and the quark-antiquark
potential are studied through the Wilson 
Polyakov loop estimations. 
In section 4, the possible bound state for the mesons and baryons are
discussed. The summary is given in the final section.

\section{Background geometry and D7 brane embedding}

We solve the equations of motion for 10d IIB model retaining the dilaton
$\Phi$, axion $\chi$ and selfdual five form field strength $F_{(5)}$.
Under the Freund-Rubin
ansatz for $F_{(5)}$, 
$F_{\mu_1\cdots\mu_5}=-\sqrt{\Lambda}/2~\epsilon_{\mu_1\cdots\mu_5}$ 
\cite{KS2,LT}, and for the 10d metric as $M_5\times S^5$ or
$ds^2=g_{MN}dx^Mdx^N+g_{ij}dx^idx^j$, we can find the solution.
The five dimensional $M_5$ part of the
solution is obtained by solving the following 5d reduced action,
\beq
 S={1\over 2\kappa^2}\int d^5x\sqrt{-g}\left(R+3\Lambda-
{1\over 2}(\partial \Phi)^2+{1\over 2}e^{2\Phi}(\partial \chi)^2
\right), \label{5d-action}
\eeq
which is written 
in the string frame and taking $\alpha'=g_s=1$. Then, we obtain,
(see the Appendix A)
$$ 
ds^2_{10}=G_{MN}dX^{M}dX^{N} ~~~~~~~~~~~~~~\hspace{6.5cm}
$$ 
\beq
=e^{\Phi/2}
\left\{
{r^2 \over R^2}A^2\left(-dt^2+a(t)^2(dx^i)^2\right)+
\frac{R^2}{r^2} dr^2+R^2 d\Omega_5^2 \right\} \ , 
\label{finite-c-sol}
\eeq 
\beq
e^\Phi= 1+\frac{q}{r^4}{1-(r_0/r)^2/3\over (1-(r_0/r)^2)^3} \ , 
\quad \chi=-e^{-\Phi}+\chi_0 \ ,
\label{dilaton}
\eeq
\beq
  A=1-({r_0\over r})^2, \quad a(t)=e^{2{r_0\over R^2} t}
\eeq
where $M,~N=0\sim 9$ and
$R=\sqrt{\Lambda}/2=(4 \pi N)^{1/4}$. As for the integration constants,
$r_0$ denotes the
horizon point, and it is related to the 4d cosmological constant $\lambda$
as
\beq
  \lambda=4{r_0^2\over R^4}.
\eeq
And $q$ is a constant which corresponds to the VEV of 
gauge fields
condensate~\cite{GY}. 
And other field configurations are set to be zero here.
We notice that there is no singularity at the horizon $r_0$ in this case.
This situation is similar to the black hole configuration used
for the finite temperature gauge theory.

\vspace{.3cm}
In the present configuration, the four dimensional boundary represents the inflational
universe characterized by the 4d cosmological constant $\lambda$. So the above
bulk-solution describes the gauge theory in the inflation universe.
Firstly, we examine the properties of the flavor quarks in this gauge theory,
and it is obtained by solving the embedding problem of the D7 
brane in this background.

\vspace{.3cm}
{\bf D7 brane embedding and phase transition:}~
The D7 brane is embedded as follows. 
The extra six dimensional part of the above metric (\ref{finite-c-sol})
is rewritten as,
\beq
 \frac{R^2}{r^2} dr^2+R^2 d\Omega_5^2
 =\frac{R^2}{r^2}\left(d\rho^2+\rho^2d\Omega_3^2+(dX^8)^2+(dX^9)^2
\right)\ ,
\eeq
where $r^2=\rho^2+(X^8)^2+(X^9)^2$.
And we obtain the induced metric for D7 brane,
$$ 
ds^2_8=e^{\Phi/2}
\left\{
{r^2 \over R^2}A^2\left(-dt^2+a(t)^2(dx^i)^2\right)+\right.
\hspace{3cm}
$$
\beq
\left.\frac{R^2}{r^2}\left((1+(\partial_{\rho}w)^2)d\rho^2+\rho^2d\Omega_3^2\right)
 \right\} \ , 
\label{D7-metric}
\eeq
where we set as $X^9=0$ and $X^8=w(\rho)$
without loss of generality due to the rotational invariance in
$X^8-X^9$ plane. Then the embedding problem is reduced to obtain the solution
for the profile function $w(\rho)$, and it is performed as follows.

\vspace{.3cm}
The brane action for the D7-probe is given as
$$ 
S_{\rm D7}= -\tau_7 \int d^8\xi \left(e^{-\Phi}
    \sqrt{-\det\left({\cal G}_{ab}+2\pi\alpha' F_{ab}\right)}
      -{1\over 8!}\epsilon^{i_1\cdots i_8}A_{i_1\cdots i_8}\right)
$$
\beq
   +\frac{(2\pi\alpha')^2}{2} \tau_7\int P[C^{(4)}] \wedge F \wedge F\ ,
\label{D7-action}
\eeq
where $F_{ab}=\partial_aA_b-\partial_bA_a$.
${\cal G}_{ab}= \partial_{\xi^a} X^M\partial_{\xi^b} X^N G_{MN}~(a,~b=0\sim 7)$
and $\tau_7=[(2\pi)^7g_s~\alpha'~^4]^{-1}$ represent the induced metric and
the tension of D7 brane respectively.
And $P[C^{(4)}]$ denotes the pullback of a bulk four form potential,
\beq
C^{(4)} = 
\left(\frac{r^4}{R^4} d x^0\wedge d x^1\wedge
d x^2 \wedge d x^3 \right)\ .
\label{c4}
\eeq
The eight form potential $A_{i_1\cdots i_8}$,
which is the Hodge dual to the axion, couples to the
D7 brane minimally. In terms of the Hodge dual field strength,
$F_{(9)}=dA_{(8)}$ \cite{GGP}, the potential $A_{(8)}$ is obtained. 

When the gauge potentials $A_a$ are neglected, the action is abbreviated as
\beq
S_{\rm D7}= -\tau_7 \int d^8\xi \left(e^{-\Phi} \sqrt{\cal G} 
      -{1\over 8!}\epsilon^{i_1\cdots i_8}A_{i_1\cdots i_8}\right) \ ,
\label{D7-action2}
\eeq
and by taking the canonical gauge, we arrive at the following D7 brane
action,
\beq
S_{\rm D7} =-\tau_7~\int d^8\xi  \sqrt{\epsilon_3}\rho^3 a(t)^3
\left(A^4
   e^{\Phi}\sqrt{ 1 + (w')^2 }-C_8 \right)
\ ,
\label{D7-action-2}
\eeq
\beq
 C_8(r)=\int^rdr'~A^4(r')\partial_{r'}\left({\rm exp}({\Phi(r')})\right)
={q\over r^4} . \label{A8}
\eeq
Then, finally, the equation of motion for $w(\rho)$ is obtained as, 
\bea
   {w\over \rho+w~w'}
   \left[\Phi'-\sqrt{1+(w')^2}(\Phi+4\log A)'~\right]
\nonumber\\
   +{1\over \sqrt{1+(w')^2}}
\left[w'\left({3\over \rho}+(\Phi+4\log A)'\right)
+ {w''\over 1+(w')^2}\right]
   &=&0 ,
  \label{qeq}
\eea
where prime denotes the derivative with respect to $\rho$. 
By solving this
equation we find the profile of the embedded D7 brane and then 
we find simultaneously the
quark properties, the quark mass $m_q$ and the chiral condensate 
$\la\bar{\Psi}\Psi\ra$, where $\Psi$ denotes the quark field. This point
is shown below through various explicit solutions.

\vspace{.3cm}
{\bf Trivial Solution and Chiral Symmetry:}~~ Firstly we consider
the asymptotic solution of $w$ for large $\rho$. 
It is obtained usually as the following form
\beq
   w(\rho) \sim m_q+{c\over \rho^2} ,  \label{asym}
\eeq
where $m_q$ and $c$ are interpreted from the gauge/gravity 
correspondence as
the current quark mass and the chiral condensate, 
respectively, since the field $\phi^8$ is corresponding to 
the conformal dimension three operators of the gauge theory. 
General solution of (\ref{qeq}) is
characterized by these two arbitrary parameters, $m_q$ and $c$, since 
Eq.~(\ref{qeq}) is the second order differential equation of $w$ although
it is highly non-linear. However, for a fixed $m_q$, we find one solution 
which is meaningful from the point of view of holography \cite{GY}. 
In other words, $c$ is determined when $m_q$ is fixed.  
This can be interpreted in the gauge theory as that 
$-c=\la\bar{\Psi}\Psi\ra$ is determined by a theory
with a parameter $m_q$, the quark mass.
Then the solutions for $w$ are characterized only by the quark mass $m_q$, 
and the vev of quark condensate is determined uniquely by $m_q$.

\vspace{.3cm}
However we must be careful in using the above asymptotic form (\ref{asym})
in solving the equation (\ref{qeq}) since
the form (\ref{asym}) is useful only for the case that CFT is realized
on the boundary $r\sim \rho \to \infty$.
In the present case, the geometry of the
4d boundary is dS${}_4$ and the conformal symmetry is broken there, then
the form (\ref{asym}) is not useful in
getting the solutions of (\ref{qeq}). 

Actually, we find a solution of $m_q=0$, and arbitrary $c$
when we solve the Eq.~(\ref{qeq}) by expanding it in terms of 
the power series of $1/\rho^2$ with the asymptotic form of (\ref{asym}).
Then, we find that the meaningful solution of 
$m_q=0$ is nothing but the trivial solution, $w=0$. 
The result is however non-trivial and important, 
because it implies that the chiral symmetry is preserved in the dS${}_4$. 

\begin{figure}
 \begin{center}
\begin{minipage}{0.95\hsize}
 \begin{minipage}{0.95\hsize}
  \hspace{3ex}
   \includegraphics[width=100mm]{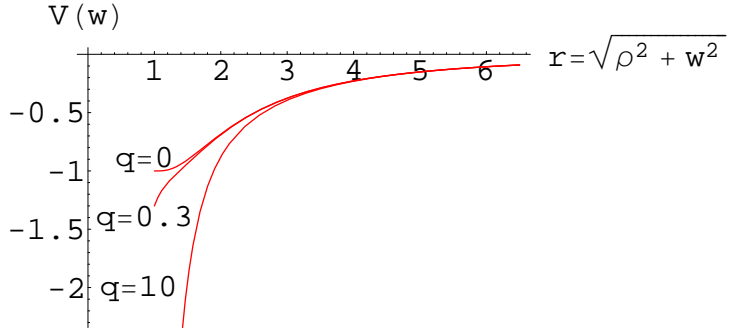}
  \vspace{-3ex}
\caption{Potential of $w$, $V(w)=A^4e^{\Phi}+C_8$, for $q=0, ~0.3$ and 10 with $R=r_0=1.0$.
\label{wq13fig}}
 \end{minipage}
 \end{minipage}
\end{center}
\end{figure}

This point is understood from the fact
that the attractive force is working between the D3-branes at the horizon
and the D7 brane at $X^8=w$. The force between them is obtained from the 
potential of $w$, which is 
obtained from the D7 action (\ref{D7-action-2})
by setting $w'=0$ and remembering $r^2=\rho^2+w^2$ as follows,
\beq
 V(w)=\tau_7\left(A^4~e^{\Phi}-C_8\right).
\eeq
It represents the potential of $w$ at fixed $\rho$, and typical examples
for different $q$s are shown in the Fig.~\ref{wq13fig}.

Then the force $F$ between
D3 and D7 branes is obtained
\beq
 F=-{\partial V\over\partial w}=-8\tau_7{w~r_0^2\over r^4}~A^3~e^{\Phi}~<0,
\label{force}
\eeq
and we can see that the force is attractive at any point of $\rho$.

Then we can understand that the
$c$ must be negative for any solution of $w$. As a result, in the case of 
$m_q=0$, we arrive at the trivial solution, $m_q=c=0$ or $w=0$ for 
non-zero $\lambda$. This situation is similar to the case of finite temperature
gauge theory, but we find several different features through the 
non-trivial solutions between
the finite temperature and finite cosmological constant case.

\begin{figure}
 \begin{minipage}{1.0\hsize}
 \begin{minipage}{0.95\hsize}
  \hspace{-3ex}
   \includegraphics[width=90mm]{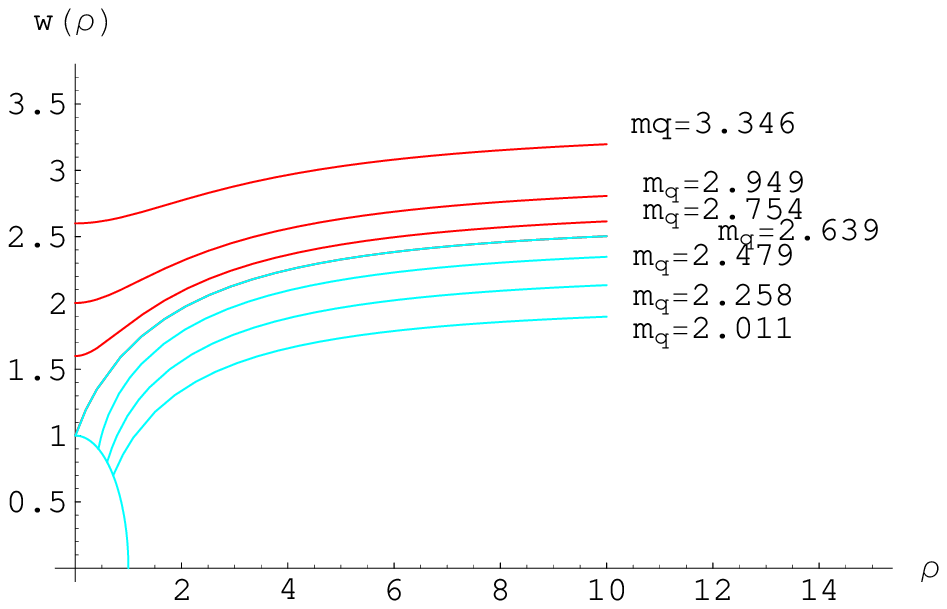}
   \includegraphics[width=80mm]{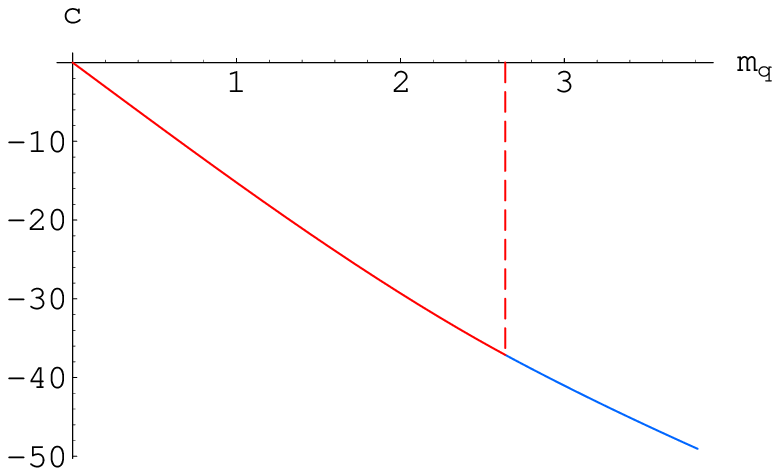}
  \vspace{0.3ex}
 \caption{Embedding solutions $w$ and the chiral condensate $c$ 
for $q=0$ and $r_0=R=1.0$. 
For $c$, the cutoff is taken at $\rho=100$, and
the dashed line shows the transition point from group 
(a) to (b) at $m_q=2.639$.
 \label{wq0fig}}
 \end{minipage}
 \end{minipage}
\end{figure}

\begin{figure}
 \begin{minipage}{1.0\hsize}
 \begin{minipage}{0.95\hsize}
  \hspace{-3ex}
   \includegraphics[width=90mm]{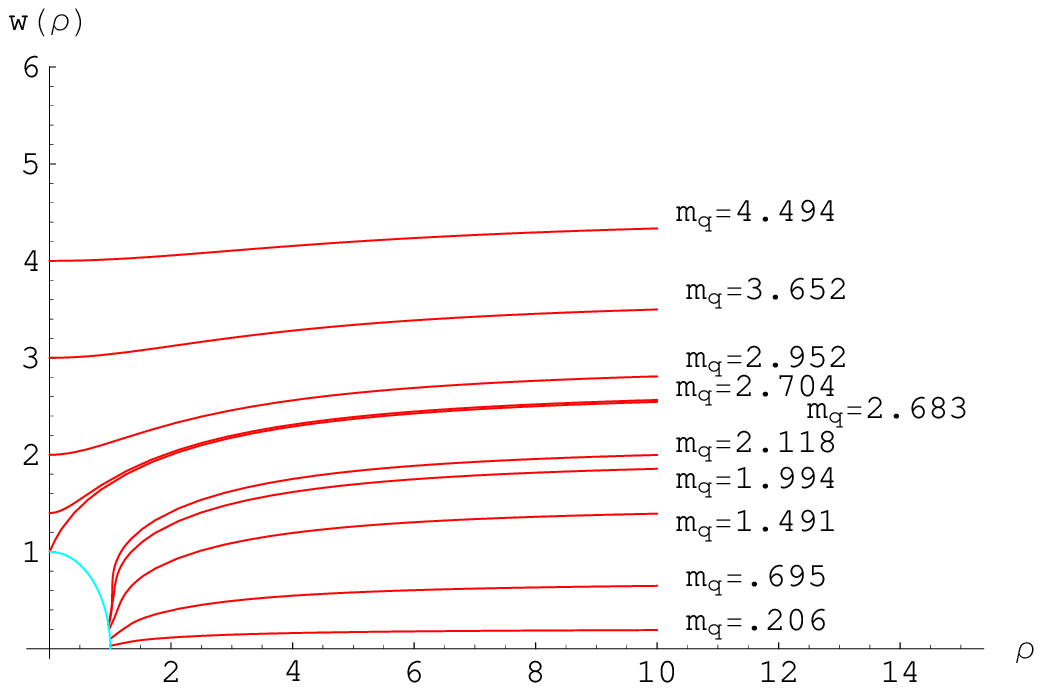}
   \includegraphics[width=80mm]{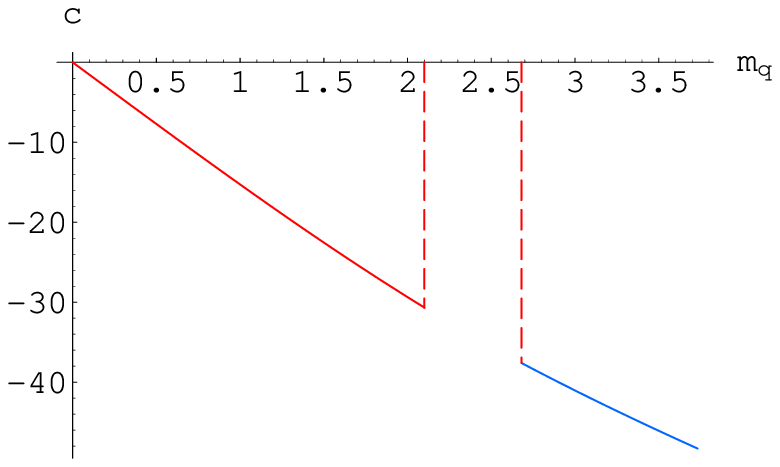}
  \vspace{0.1ex}
\caption{For $q=0.3$ and other parameter settings are the same with
the above Fig.~\ref{wq0fig}. The dashed lines in the right 
figure show the two critical points, $m_q=2.683$ and $=m_q=2.118$.
\label{wq1fig}}
 \end{minipage}
 \end{minipage}
\end{figure}

\begin{figure}
 \begin{minipage}{1.\hsize}
 \begin{minipage}{0.95\hsize}
  \hspace{-3ex}
   \includegraphics[width=90mm]{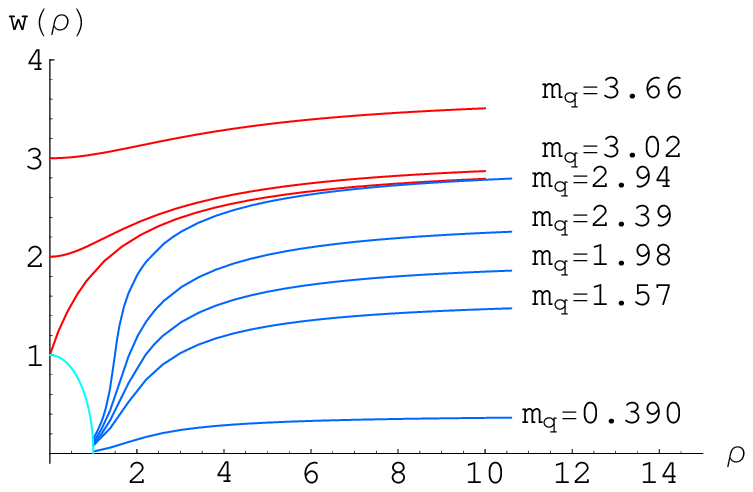}
   \includegraphics[width=80mm]{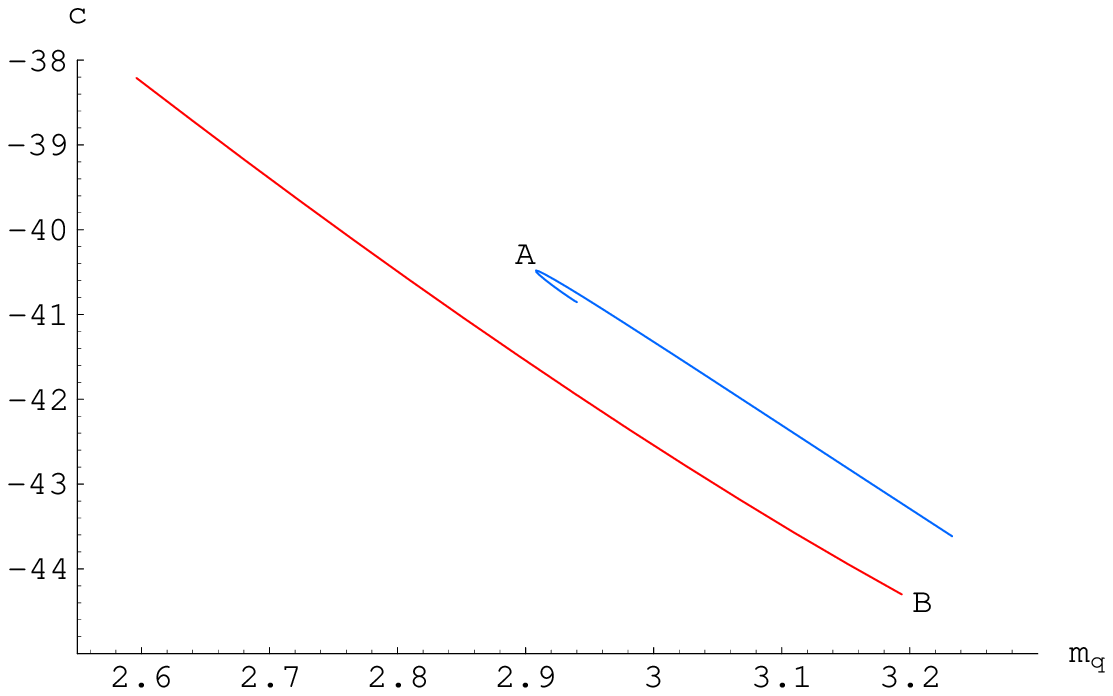}
  \vspace{.2ex}
\caption{For $q=10$ and other parameter settings are the same with
the above Fig.~\ref{wq0fig}. 
\label{wq12fig}}
 \end{minipage}
 \end{minipage}
\end{figure}

\vspace{.3cm}
{\bf Non-trivial Solutions and Phase Transition:}~
However, the trivial solution is a part of solutions. In fact,
we could find many non-trivial numerical solutions, whose behavior seems to be 
consistent with the asymptotic form of
(\ref{asym}) at a glance. 
Some of them are shown
in the Fig.~\ref{wq0fig}-\ref{wq12fig} for different $q$s. 
In order to obtain these solutions in terms of the
power series expansions, we must
add correction terms of $\log(\rho)$ to the chiral condensate $c$ and to the 
other
coefficients of the power series of $1/\rho^2$.
This improvement is naturally expected since the VEV, 
$c=-\la\bar{\Psi}\Psi\ra$, would receive this type of loop corrections due to
the conformal non-invariance of the gauge theory
in the present background dS${}_4$.

Then, instead of (\ref{asym}),
we find the following asymptotic form 
\beq
  w(\rho) \sim m_q+{c_0-4m_q r_0^2\log(\rho)\over \rho^2} ,  \label{asym2}
\eeq
where $m_q$ and $c_0$ are arbitrary at this stage. However, $c_0$ would be 
determined by $m_q$ and other parameters of the theory for the meaningful 
solution as stated above. 
And the solutions of $w(\rho)$ and $c$
for various $m_q$ are shown by separating the solutions
to three groups depending on the value of $q$.

As for the $c$, in the present case, it
depends on $\log(\rho)$ and diverges at
$\rho=\infty$. So we need an appropriate subtraction or a 
renormalization for this quantity \cite{H}. However, we estimate it 
as 
\beq
c=c_0-4m_q r_0^2\log(\rho_{\rm cutoff})
\eeq
here by introducing an appropriate 
cutoff $\rho_{\rm cutoff}$ for the simplicity.

\vspace{.3cm}
In any case, there is a horizon at $\rho^2+w(\rho)^2=r_0^2$
in the $\rho -w$ plane, and the embedded solutions are separated to two
categories by the infrared end point of $w(\rho)$. In the first group, 
it is above the horizon ({\bf group (a)}), and for the second, it is on the 
horizon ({\bf group (b)}). This situation is also seen
in the case of the finite temperature gauge theory. 

The solutions are shown in the Fig.~\ref{wq0fig}-\ref{wq12fig} for
different $q$s and we find that all the solutions of
$w$ decreases with decreasing $\rho$, which means 
$c\,(=c_0-4m_q r_0^2\log(\rho)) \leq 0$,
for any solution. This is, as mentioned above, due to the 
attractive force between 
D3 and D7 branes, and $c=0$ for $m_q=0$ as expected. In the Fig.~\ref{wq13fig},
we show the potential for the three different $q$ cases for our understanding
of this point.
In other words, the chiral symmetry is preserved for $m_q=0$ in any case. 
The numerical results are summarized as follows.

\vspace{.3cm}
{\bf (i) $q=0$:}~~Firstly we consider the case of $q=0$ or zero 
gauge field condensate, shown in the Fig.~\ref{wq0fig},
$\la F_{\mu\nu}^2\ra=0$. In this state, the quarks are not confined even in the
limit of $\lambda=0$. 
The second point to be noticed is that we could not find any jump of solutions
when it changes from the group (a) to the group (b), and $c$ changes smoothly
at the transition point of $m_q$.

\vspace{.5cm}
For finite $q$, the solutions for the group (b) are strongly restricted.
Namely, the end point value of $w_0\equiv w(\rho_0)$, 
where $\rho_0$ denote the value of $\rho$ on the horizon, is constrained
as (see Appendix B)
\beq
  w_0 < {r_0\over \sqrt{10}}   \label{constraint}
\eeq
and the maximum of $w_0$ is realized in the limit $q\to 0$ but $q\neq 0$.
We find solutions of $m_q<0.96$ (in Gev) for $q=10^{-7}$ and $r_0=1$. On the
other hand, the solution for $q=0$ and $r_0=1$ has $m_q >2.4$, so there is a
gap for the quark mass of groups (a) and (b). This gap decreases with 
increasing $q$, and we summarize the numerical results for $q>0$ as follows.

\vspace{.5cm}
{\bf (ii) Small $q (<2)$:}~~
We show in the Fig.~\ref{wq1fig} the solutions for non-zero but small 
$q$, namely at $q=0.3$. The $\rho$ dependence of the solutions are 
similar to the $q=0$ case,
but there is a forbidden region of solutions for $2.118<m_q<2.683$ (in GeV)
between the two solution groups of (a) and (b).
This phenomena is characteristic to the present case of dS${}_4$, and it
has not been seen in the finite temperature case. This gap gradually
decreses with increasing $q$ and disappears for $q>2$ as shown below. 

\vspace{.3cm}
{\bf (iii) Large $q (>2)$:}~~
In the Fig.~\ref{wq12fig}, we show the solutions for large $q(=10)$ case.
In this case, the behavior of the solutions is similar to
the case of the finite temperature gauge theory. Namely, the forbidden band of
the quark mass or a gap of $w(\infty)$ at the boundary seen in (ii) case
disappears, but the gap at the infrared end point of $w$ still remains and a
phase transition as seen in the finite temperature case is expected
when the solution translates from group (a) to the group
(b). This is seen as a gap of the infrared end point $w$ for the solution of 
common $m_q$ at the transition point. In the present example,
it is realized at $m_q=2.94(\equiv m_c)$.
This transition is also seen through the chiral condensate $c$, which
is actually seen from the Fig.~\ref{wq12fig} as a jump of $c$ at $m_q=m_c$.
In the figure showing $c$, the upper one shows the one of solution group (a)
and it begins with a gap at 
$m_q\sim 2.9$. And there is a narrow double valued region near the point A. 
In the multiple valued region of $c$, we choose the value of $c$, whose
D7 energy is the minimum, as the plausible solution. The lower curve is 
obtained from the solutions of group (b) and it ends at $m_q\sim 3.2$, 
the point B.
The similar gap is seen also in the mass spectra
of mesons as seen below.

\vspace{.3cm}
Next, we study the reguralized D7 energy $E_{\rm D7}$ defined as
\bea
S_{\rm D7} &=&-\tau_7~\int d^7\xi  \sqrt{\epsilon_3}a(t)^3E_{\rm D7} \\
E_{\rm D7}&=&\int_{\rho_{\rm min}}^{\infty}d\rho~ \rho^3 \left(A^4
   e^{\Phi}\sqrt{ 1 + (w')^2 }-C_8 \right)
\ ,
\label{D7-energy}
\eea
in order to see the order of the phase 
transition which is clearly expected in the above calculation of case (iii)
for $q (>2)$. The regularization for the D7 energy for the finite $m_q$ 
solution
is usually performed by subtructing the one of $m_q=0$ \cite{KMMW}, and
the regularized energy $E_{\rm D7}^{\rm reg}(m_q)$ for the mass $m_q$ is
given as
\bea
E_{\rm D7}^{\rm reg}(m_q)&=& E_{\rm D7}(m_q)-E_{\rm D7}(0) \\
  &=&\int_{\rho_{\rm min}}^{\rho_{\rm mutch}}d\rho~ \rho^3 
     \left(A^4e^{\Phi}\sqrt{ 1 + (w')^2 }-{q\over r^4}\right)\Big|_{w(m_q)} 
\nonumber \\
 && -\int_{r_0}^{\rho_{\rm mutch}}d\rho~ \rho^3 
     \left(A^4e^{\Phi}-{q\over \rho^4}\right)\Big|_{w=0} 
  +\int_{\rho_{\rm mutch}}^{\infty}d\rho~\Delta L_{\rm D7} \, , 
\label{reg-energy}
\eea
where $\rho_{\rm mutch}$ represents the position where $w$ is approximated well
by its asymptotic form (\ref{asym2}).
In the present case, however, we need an extra subtraction term due to the 
cosmological constant $r_0$. 
\begin{figure}
 \begin{minipage}{1.\hsize}
 \begin{minipage}{0.95\hsize}
  \hspace{-3ex}
   \includegraphics[width=90mm]{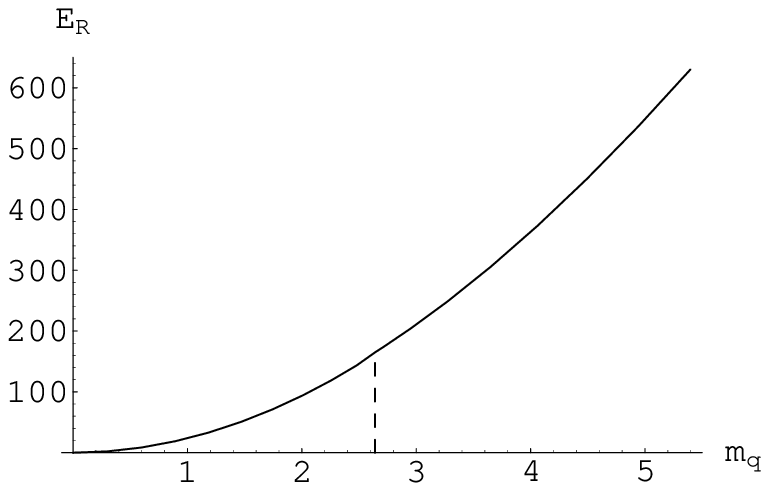}
  \includegraphics[width=80mm]{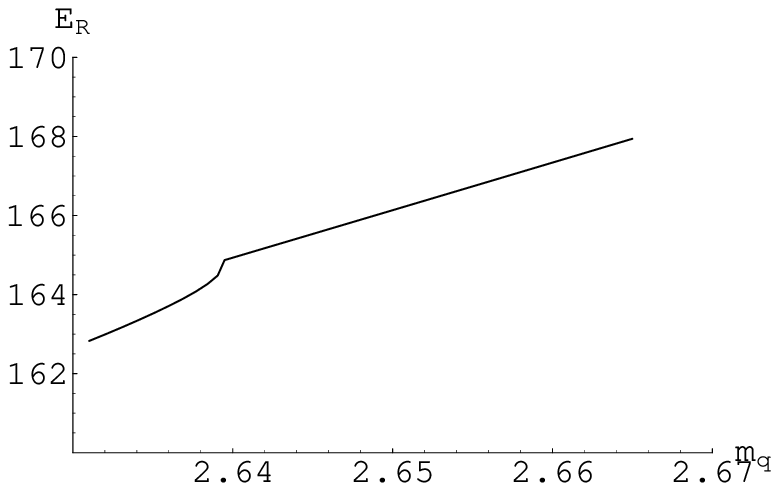}
  \vspace{.2ex}
\caption{The regularized energy
for $q=0$, 
and other parameter settings are the same with
the above Fig.~\ref{wq0fig}. \label{wq15fig}}
 \end{minipage}
 \end{minipage}
\end{figure}
\begin{figure}
 \begin{minipage}{1.\hsize}
 \begin{minipage}{0.95\hsize}
  \hspace{-3ex}
   \includegraphics[width=80mm]{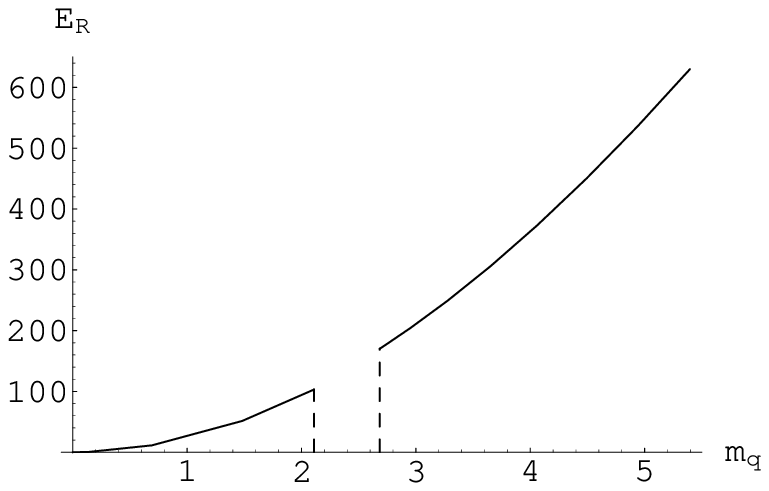}
  \vspace{.2ex}
\caption{The regularized energy
for  $q=0.3$, 
and other parameter settings are the same with
the above Fig.~\ref{wq0fig}. \label{wq16fig}}
 \end{minipage}
 \end{minipage}
\end{figure}
\begin{figure}
 \begin{minipage}{1.\hsize}
 \begin{minipage}{0.95\hsize}
  \hspace{-3ex}
   \includegraphics[width=90mm]{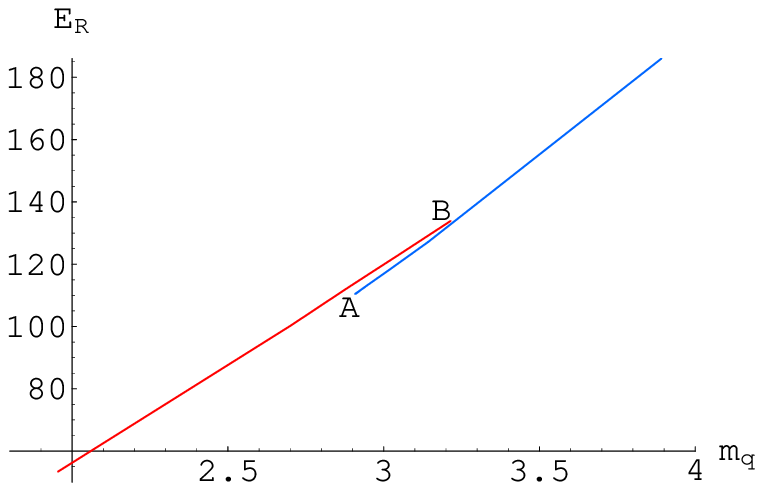}
  \includegraphics[width=80mm]{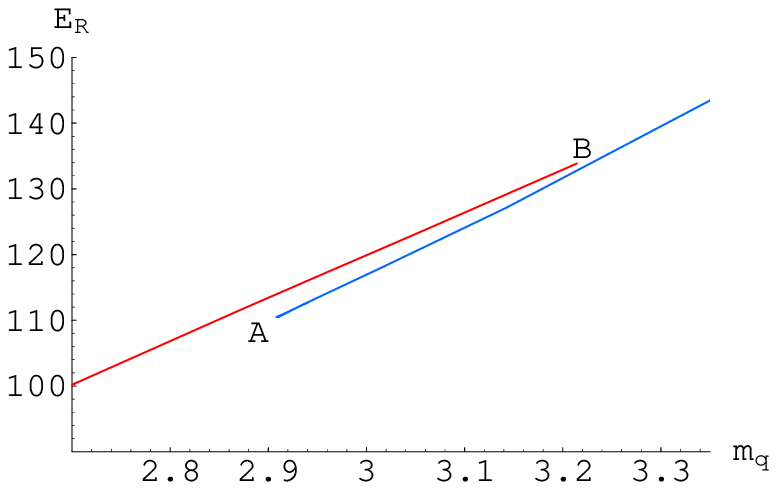}
  \vspace{.2ex}
\caption{The regularized energy
for $q=10$, and other parameter settings are the same with
the above Fig.~\ref{wq0fig}. The points A and B are the same points with the onein  Fig.~\ref{wq12fig}. \label{wq14fig}}
 \end{minipage}
 \end{minipage}
\end{figure}
In fact, in the large
$\rho$ region, the integrand $\Delta L_{\rm D7}$ is expanded by 
using asymptotic form of $w(\rho)$ as follows
\beq
 \Delta L_{\rm D7}=4{m_q^2r_0^2\over \rho}+O(\rho^{-3})
\eeq
so we subtract here the first term in the estimation of 
$E_{\rm D7}^{\rm reg}(m_q)$. Then we obtain a small correction from the third
term of Eq.(\ref{reg-energy}).

\vspace{.3cm}
The numerical results are shown in the Fig.\ref{wq15fig}-\ref{wq14fig}.
In the case of $q=0$, we can see a small cusp-like structure at the transition
point between the two types of solutions (a) and (b). 
This might imply some
kind of phase transition. On the other hand, 
for $q=10$, we find definite energy gap between the solutions of two groups 
at the same value of $m_q$ in their overlap region. We then expect
a first order phase transition from a solution of group (b) to the one of 
group (a) at an appropriate point of $m_q$. The details of these transitions
observed here will be given in the future.


\section{Quark-antiquark potential and quark confinement}

We study a gravity description of 
quark-antiquark potentials in order to study the quark confinement.
Before giving the calculation, we briefly review how quark-antiquark 
potentials described
in the context of the gauge/gravity correspondence. 

Consider the Wilson-Polyakov loop,  
$
   W={(1/N)} \textrm{Tr} P e^{i\int A_0 dt} 
$, 
in $SU(N)$ gauge theory, then
the quark-antiquark potential $V_{q\bar{q}}$ is derived from
the expectation value of a parallel Wilson-Polyakov loop as
$ 
   \langle W\rangle \sim e^{-V_{q\bar{q}}\int dt} 
$. 
Meanwhile, from the gravity side the expectation value is represented as
\beq
    \langle W\rangle  \sim e^{-S} , \label{wstr}
\eeq
in terms of the Nambu-Goto action 
\beq
   S=- \frac{1}{2 \pi \alpha'} 
\int d\tau d\sigma \sqrt{-\textrm{det}\, h_{ab}} , 
\eeq
with the induced metric
$ 
    h_{ab}=G_{\mu\nu}\partial_a X^{\mu}\partial_b X^{\nu} ,
$ 
where $X^{\mu}(\tau,\sigma)$ denotes the string coordinate.

\vspace{.3cm}
Then the quark-antiquark potential can be calculated
by setting the configurations of string coordinates under the background
geometry given above.
To fix the static string configurations 
of a pair of quark  and anti-quark,
we choose $X^0=t=\tau$ and decompose 
the other nine string coordinates into components 
parallel and perpendicular to the D3-branes:
\beq
 \mathbf{X} = (\mathbf{X}_{||}, r, r \Omega_5) .
\eeq
Then the Nambu-Goto Lagrangian in the background (\ref{dilaton}) 
becomes
\beq
   L_{\textrm{\scriptsize NG}}=-{1 \over 2 \pi \alpha'}\int d\sigma ~e^{\Phi/2}
   \sqrt{A(r)^2r'{}^2+r^2A(r)^2\Omega_5{}'{}^2
        +\left({r\over R}\right)^4 A(r)^4a(t)^2 \mathbf{X}_{||}{}'{}^2} ,
 \label{ng}
\eeq
where the prime denotes a derivative with respect to $\sigma$.
The test string has two possible configurations:
(i) a pair of parallel string, which connects horizon and the D7 brane,
and (ii) a U-shaped string whose two end-points are on the D7 brane.

\vspace{.3cm}
{\bf Effective quark mass and confinement:}
Firstly consider the configuration (i) of parallel two strings,
which have no correlation each other.
The total energy is then two times of one effective quark mass,
$\tilde{m}_q$. As mentioned
above, it is given by a string configuration which stretches 
between  $r_0$ and the maximum $r_{\textrm{\scriptsize max}}$,
so we can take as
\beq
   r=\sigma,~~~ \mathbf{X}_{||}=\textrm{constant},~~~
   \Omega_5=\textrm{constant}. 
\label{para}
\eeq
Then $\tilde{m}_q$ is obtained by substituting 
(\ref{para}) into (\ref{ng}) as follows,
\beq
   E=
   {1\over \pi \alpha'}\int_{r_0}^{r_{\textrm{\scriptsize max}}}
     dr ~e^{\Phi/2}A(r) = 2\tilde{m}_q \ , \label{dynamicalmass}
\eeq
where $r_{\rm rmax}$ denotes the position of the D7 brane.

\begin{figure}[htbp]
\vspace{.3cm}
\begin{center}
\includegraphics[width=11.5cm]{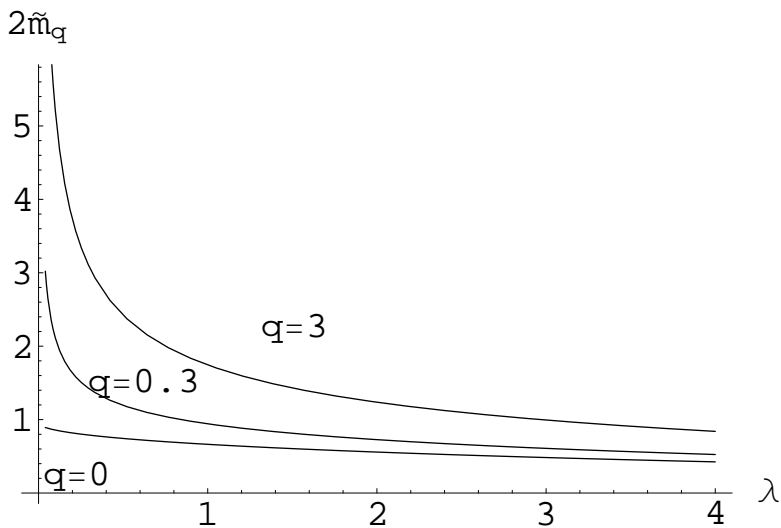}

\caption{$2\tilde{m_{q}}$ are shown for $R=1\,(\mbox{GeV}^{-1}),r_{max}=3$ 
  and $\alpha'=1\,(\mbox{GeV}^{-2})$. The three curves are corresponding 
  to the case of $q$=0, 0.3 and 3 respectively.
\label{el-0}}
\end{center}
\end{figure}

\vspace{.3cm}
The integrand $e^{\Phi/2}A(r)$ is diverges at $r=r_0$ as $1/\sqrt{r-r_0}$,
but we find that the contribution of this part to the integration vanishes,
\beq
  \int_{r_0}dr~ {1\over\sqrt{r-r_0}}=2\sqrt{r-r_0}\left.\right |_{r_0}=0\ .
\eeq
Then we find $\tilde{m_{q}} < \infty$ for finite $r_0$ or $\lambda$. This means
that the quark is not confined in this case since single quark could exist. 
On the other hand, we find $\tilde{m}_q$ diverges for $r_0=0$ then the quark 
confinement. 

While, in the above discussion, $q\neq 0$ is essential, we find for $q=0$
\beq
  2\tilde{m_{q}}={1\over \pi \alpha'}{(r_{\rm max}-r_0)^2\over r_{\rm max}}\ .
\eeq
Then the quark is not confined in this case even if $\lambda=0$.

As for the cosmological constant dependence on $\tilde{m}_q$, the numerical
calculations are shown in Fig.~\ref{el-0}. From this Fig., we find that
(i) $\tilde{m}_q$ increases with $q$ at any point of $\lambda$, and (ii)
it decreases with increasing $\lambda$ for any $q$.

\begin{figure}[htbp]
\vspace{.1cm}
\begin{center}
\includegraphics[width=7.cm]{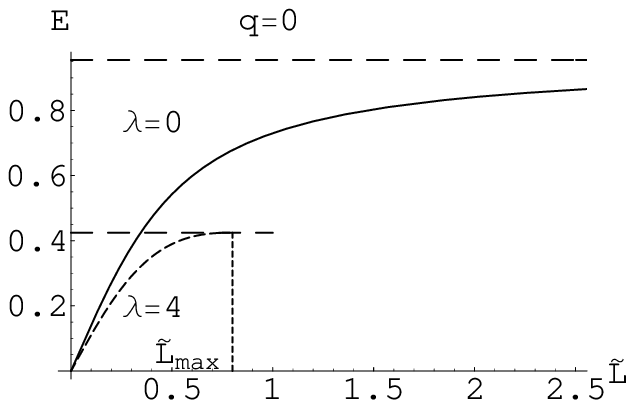}
\includegraphics[width=7.cm]{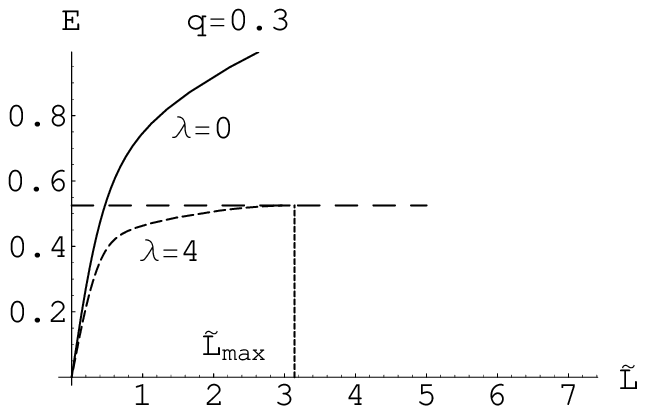}
\caption{{\small
Plots of $E$ vs $\tilde{L}$ at $q=0$ (the left figure)
and $=0.3$ (the right figure) $(\mbox{GeV}^{-4})$
for $R = 1\, (\mbox{GeV}^{-1})$, 
$r_{\textrm{\scriptsize max}}=3\, (\mbox{GeV}^{-1})$ 
and $\alpha'=1\, (\mbox{GeV}^{-2})$.
The solid and dashed curves represent the case of 
$\lambda=0 $ and $\lambda=4\,(\mbox{GeV})$, respectively. 
The vertical solid and dashed lines represent the energy of two 
parallel straight strings.}
\label{el-01}}
\end{center}
\end{figure}

\vspace{.5cm}
\noindent {\bf U-shaped Wilson-Loop:}
We now turn to the U-shaped configuration,
\beq
    \mathbf{X}_{||}=(\sigma,0,0),~~~
    \Omega_5=\textrm{constant} .
    \label{ushape}
\eeq
The equation of motion derived from the Lagrangian (\ref{ng}) with
the configuration (\ref{ushape})
are solved by
\beq
     e^{\Phi/2}{1\over \sqrt{(r/R)^4 A^2(r)a^2(t)+(dr/d\sigma)^2}}
    \left({r\over R}\right)^4 A^3(r)a^2(t)= \textrm{constant} .
\eeq
The midpoint $r_{min}$ of the string is determined by $dr/d\sigma|_{r=r_{min}}=0$.
Then the distance and the total energy of the quark and anti-quark
are given by
\bea
  &&  L=2R^2 \int_{r_{min}}^{r_{\textrm{\scriptsize max}}} dr~
      {1\over r^2 A(r)a(t)
        \sqrt{e^{\Phi(r)}r^4 A(r)^4 /
          \left(e^{\Phi(r_{min})}r_{min}^4A(r_{min})^4\right)-1}} ,
\\
  && E=
   {1\over \pi \alpha'} \int_{r_{min}}^{r_{\textrm{\scriptsize max}}}
   {A(r)e^{\Phi(r)/2}\over 
     \sqrt{1-e^{\Phi(r_{min})}r_{min}^4 A(r_{min})^4/
             \left(e^{\Phi(r)}r^4 A(r)^4\right)}} .
\eea
Here we study the time independent distance
$\tilde{L}$ defined as  $\tilde{L}\equiv aL$ instead of $L$ given above.
The numerical results are shown in the Fig.~\ref{el-01}

Two figures show the dependence of the energy $E$
on the distance $\tilde{L}$ at the selected cosmological constant $\lambda$ and $q$. For $\lambda=0$, the well-known results are seen for $q=0$ and finite $q$;
(i) For $q=0$, $E\propto 1/L$ at large $L$ and $E\propto m_q^2~L$ at small $L$
\cite{KMMW}.
And for finite $q$ \cite{GY}, $E\propto \sqrt{q}L$ at large $L$ and $E\propto m_q^2~L$ at small $L$. The behaviors at small $L$ are the common since the same AdS limit
is realized there.

\vspace{.3cm}
For finite $\lambda$, there is a maximum value of $L~(=L_{\rm max})$ to 
find the U-shaped configuration. Namely, the U-shaped configuration
disappears for $L>L~(=L_{\rm max})$. The similar
behavior is seen also in the case of the finite temperature \cite{GSUY}. 
In this sense, the theory in dS${}_4$ is in the quark deconfinement phase
as in the finite temperature case.
However, we notice the following difference.
In the case of finite temperature, there are two possible U-shaped
string configurations at the same values of $L~ (<L_{\rm max})$, 
but in the case of finite 
cosmological constant, U-shaped string configuration is unique at a
given value of $\tilde{L} (\tilde{L}<\tilde{L}_{max}$). And at
$\tilde{L}=\tilde{L}_{max}$, the energy of this string configuration arrives
at $2\tilde{m}_q$. Then, this implies that the U-shaped string configuration
is broken for $L>L_{\rm max}$ to decay to free quark and anti-quark.
On the other hand, an unstable U-shaped string configuration is allowed
for the finite temperature case even if $E>2\tilde{m}_q$, and, just in this
energy region, the other U-shaped string configuration is formed.

\vspace{.2cm}
In our model, $\tilde{L}_{max}$ is given as,
\beq
\tilde{L}_{max}=\lim_{r_{min} \to r_0} \tilde{L}\ ,
\eeq
and for $q=0$, and we obtain 
\bea
\tilde{L}_{max}&=&\frac{4}{\sqrt{\lambda}}\lim_{x_{a} \to 1}\displaystyle\int_{x_{b}}^{x_{a}}  dx
\frac{1}{(1-x^2)}\biggl(\frac{x_a^4(1-x^2)^4}{x^4(1-x_a^2)^4}-1 \biggr)^{-\frac{1}{2}}\\ 
&\sim&
\frac{\pi}{2}\frac{1}{\sqrt{\lambda}}\,
\eea
where 
\beq
 x \equiv \frac{r_0}{r},\quad 
x_{a} \equiv \frac{r_0}{r_{min}}, \quad x_{b} \equiv \frac{r_0}{r_{max}}\ .
\eeq
While the value of $\tilde{L}_{max}$ in general depends on the lower bound
$x_{b}$, however 
the approximate value seems to be independent
of it. This is because of the fact that the integration is approximated
by the value near the upper bound, $x_{a}$, of the integration. And we can show
the $x_{b}$ independence of $\tilde{L}_{max}$ directly by 
\beq
{\partial \tilde{L}_{max}\over \partial {x_{b}}}=0\ . \label{rm-ind}
\eeq
The numerical result is consistent with the one given in \cite{H}.

\begin{figure}[htbp]
\vspace{.3cm}
\begin{center}
 \includegraphics[width=7cm]{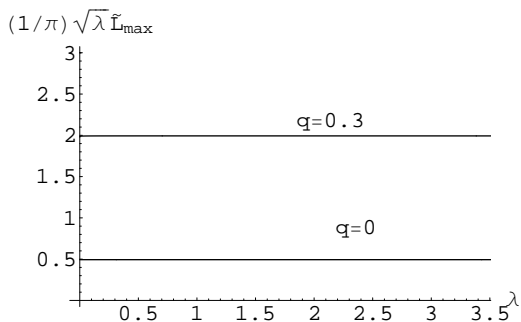}
\caption{ Functional relation between
  $\frac{\sqrt{\lambda}\tilde{L}_{max}}{\pi}$ and $\lambda$ for $q=0$
and $q\neq 0$
\label{el-3}}
\end{center}
\end{figure}

\vspace{.2cm}
As for $q\neq 0$, we get
\bea
\tilde{L}_{max}&=&
\frac{4}{\sqrt{\lambda}}
\lim_{x_{a} \to 1}
\displaystyle\int_{x_{b}}^{x_{a}}  dx
\frac{1}{(1-x^2)} \biggl(\frac{x_a^4
  \bigl(1+\tilde{q}x^4\frac{(1-\frac{1}{3}x^2)}{(1-x^2)^3} \bigr)(1-x^2)^4}{x^4\bigl(1+\tilde{q}x_{a}^4\frac{(1-\frac{1}{3}x^2)}{(1-x_{a
           }^2)^3}\bigr)(1-x_{a}^2)^4}-1 \biggr)^{-\frac{1}{2}}\\
&\sim&
\frac{2\pi}{\sqrt{\lambda}}\ ~,
\eea
where $\tilde{q}=q/r_0^4$.
In this case also, we can show the $r_{\rm max}$ independence of 
$\tilde{L}_{max}$ by (\ref{rm-ind}), but the $q$ independence of the
above result is non-trivial. This point is understood from the fact
that the dominant part of the integrant in this integration is 
independent of $q$. As a result, we
get the result which is independent of
$\tilde{q}$ and $r_{\rm max}$. But the approximate value is about four
times of the one with $q=0$. This is reduced to the 
difference of the quark-antiquark
potentials at $\lambda=0$ for $q=0$ and $q\neq 0$. For the latter case
$q\neq 0$, we obtain the linear potential at $\lambda=0$.
Their numerical values are
also plotted in the Fig.~\ref{el-3} for $r_{\rm max}=10$.

\section{Possible hadron spectrum}
As shown in the previous section, while the theory is in the quark
deconfined phase, the U-shaped string configuration exists. However,
the energy of the configuration is bounded from the above by $2\tilde{m}_q$
which is the energy of the parallel string configuration. And its value
depends on the cosmological constant, $\lambda$, and $q$. This situation
implies the existence of stable meson states, the bound state of quark and
anti-quark, when their mass is below $2\tilde{m}_q$.
We study this point by calculating the meson mass and
comparing it with $2\tilde{m}_q$. 
And we find a stable meson when its mass is smaller than $2\tilde{m}_q$
since this meson state can be expressed by the U-shaped string
configuration, namely as a bound state of quark and anti-quark.

\subsection{Meson spectra}
The meson spectrum is obtained by solving the equations of motion
of the fields on the D7 brane. According to \cite{BGN}, firstly we
consider the fluctuations of the scalar mesons which are defined as,
$$X^9=\tilde\phi^9,\quad X^8=w(\rho)+\tilde\phi^8.$$
And writing the wave functions in the following factorized form, 
$$\tilde\phi^k=\phi^k(t,x^i)\phi^k(\rho){\cal Y}_l(S^3)\qquad(k=9,8)$$
we get
the linearlized field equations for $\phi^9(\rho)$ and $\phi^8(\rho)$ 
as follows
$$
 \partial_{\rho}^2\phi^9+
    {1\over L_0}\partial_{\rho}(L_0)\partial_{\rho}\phi^9
+(1+w'~^2)
\left[({R\over r})^4{m_9^2\over A^2}-{l(l+2)\over \rho^2}-2K_{(1)}
\right]\phi^9
$$
\beq
+(1+w'~^2)^{1/2}{1\over r}{{\partial\Phi}\over{\partial r}}\phi^9=0
  \label{phi9}
\eeq
\beq
 L_0=\rho^3e^{\Phi} A^4 {1\over\sqrt{1+w'~^2}},
  \quad    K_{(1)}={1\over e^{\Phi} A^4}\partial_{r^2}(e^{\Phi} A^4)
\eeq
and
$$
 \partial_{\rho}^2\phi^8+
    {1\over L_1}\partial_{\rho}(L_1)\partial_{\rho}\phi^8
+(1+w'~^2)\left[({R\over r})^4{m_8^2\over A^2}-{l(l+2)\over \rho^2}
     -2(1+w'~^2)(K_{(1)}+2w^2K_{(2)})
\right]\phi^8
$$
$$
+(1+w'~^2)^{3/2}\left[\left(2rK_{(1)}{{\partial\Phi}\over{\partial r}}
+{{\partial^2\Phi}\over{\partial r^2}}\right){{w^2}\over{r^2}}
+{{\partial\Phi}\over{\partial r}}{{\rho^2}\over{r^3}}\right]\phi^8
$$
\beq
 =-2{1\over L_1}\partial_{\rho}(L_0 w~w'K_{(1)})\phi^8
\eeq
\beq
 L_1={L_0\over {1+w'~^2}}, \quad
   K_{(2)}= {1\over e^{\Phi} A^4}\partial_{r^2}^2(e^{\Phi} A^4).
\eeq
Where four dimensional mass $m_9$ and $m_8$ are defined by 
\beq
\ddot\phi^k(t,x^i)+3{{\dot a}\over a}\dot\phi^k(t,x^i)
+{{-\partial_i^2}\over{a^2}}\phi^k(t,x^i)=-m_k^2\phi^k(t,x^i).\qquad(k=9,8)
   \label{mass}
\eeq
In deriving the above equations of $\phi^8$ and $\phi^9$, we used
\beq
 r^2=\rho^2+(\phi^8)^2+(\phi^9)^2+w^2+2w \phi^9\ .
\eeq
But we should notice here that
the variable $r$ in the above field equations is understood as
$r^2=\rho^2+w^2$ since we are considering the linearized equations.
                                                                                \vspace{.3cm}

\begin{figure}[htbp]
\vspace{.3cm}
\begin{center}
\subfigure[] {\includegraphics[angle=0,width=0.45\textwidth]{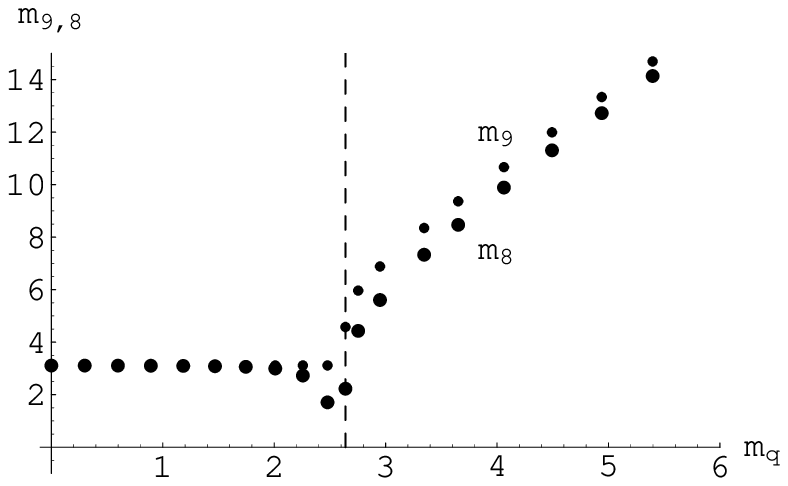}
            \label{mass-fig}}
\subfigure[] {\includegraphics[angle=0,width=0.45\textwidth]{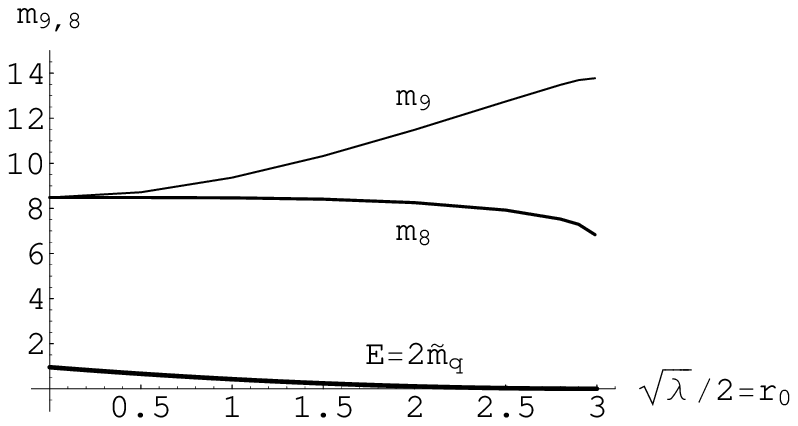}
            \label{mass2-fig}}
\subfigure[] {\includegraphics[angle=0,width=0.45\textwidth]{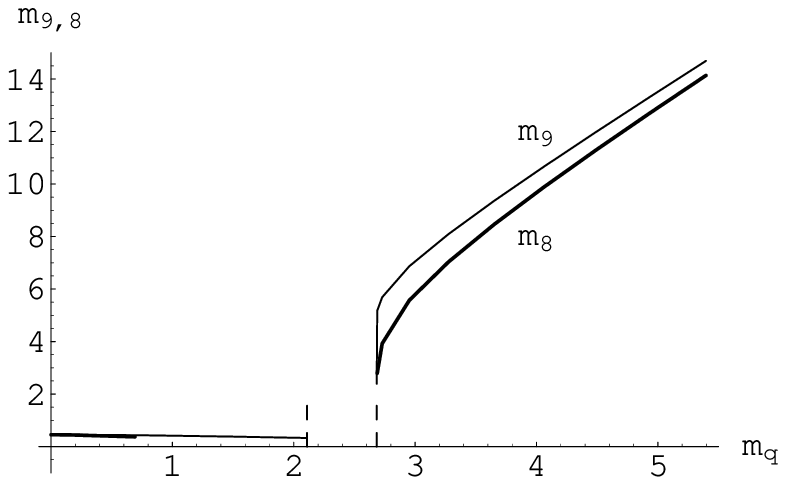}
            \label{mass3-fig}}
\subfigure[] {\includegraphics[angle=0,width=0.45\textwidth]{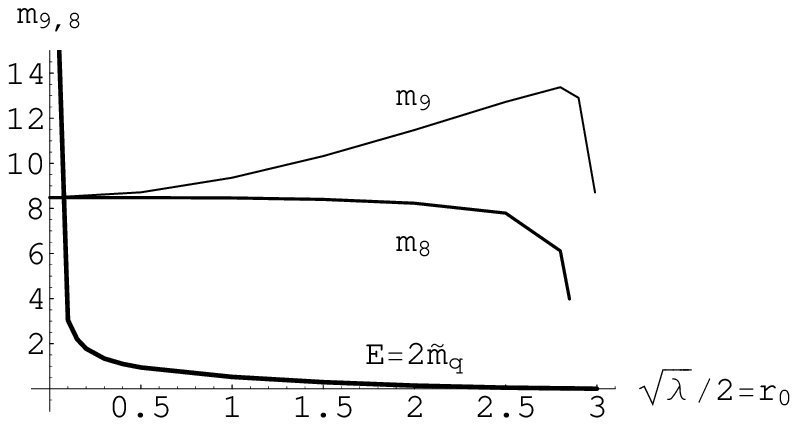}
            \label{mass4-fig}}
\subfigure[] {\includegraphics[angle=0,width=0.45\textwidth]{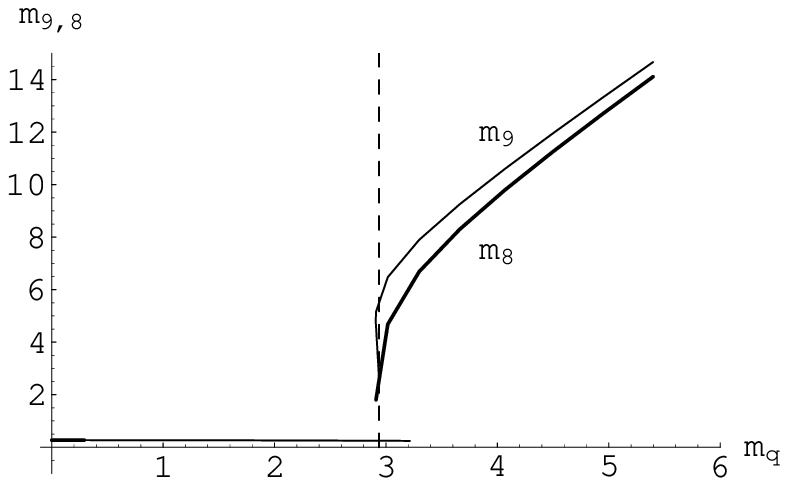}
            \label{mass5-fig}}
\subfigure[] {\includegraphics[angle=0,width=0.45\textwidth]{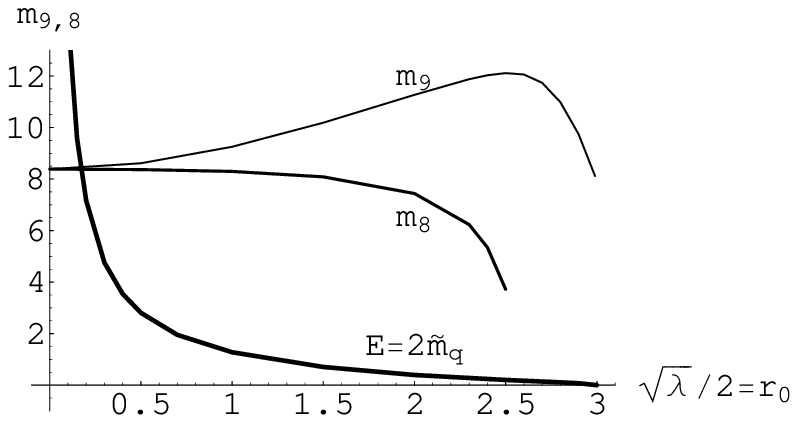}
            \label{mass6-fig}}
\caption{$m_{9,8}$ vs $m_q$ for $r_0=1.0$ and $R=1$ and
(a) $q=0$, (c) $q=0.3$, (e) $q=10$.
Small points denote $m_9$ while large points denote $m_8$ ((a)).
Thin lines denote $m_9$ while thick lines denote $m_8$ ((c), (e)).
Each vertical dashed line of the right of (a), (c) and (e) separates 
the region which touch the horizon and the region which do not touch 
the horizon.
And $m_{9,8}$ vs $r_0=\sqrt{\lambda}/2$ for $R=1$ and $w(0)=3.0$ and 
(b) $q=0$, (d) $q=0.3$, (f) $q=10$.  Thin line denotes $m_9$ while 
mid-thick lines denotes $m_8$.  Thick line denote $E=2\tilde m_q$.
}
\end{center}
\end{figure}
 
In Figs.~\ref{mass-fig}, \ref{mass3-fig}, \ref{mass5-fig},
the numerical results of the mass 
eigenvalues, $m_9$ (small point/thin line) and 
$m_8$ (large point/thick line), are plotted as functions of $m_q$.
These values are all for the nodeless solutions, i.e. for the lowest
mass state. The eigenvalues for $q=0$, Fig.~\ref{mass-fig},
show smooth variation with respect to
$m_q$ as expected from the solutions of its profile function (see
the Fig.~\ref{wq0fig}). For a non-zero but small value of $q$, $q=0.3$,
the mass eigenvalues are shown in the Fig.~\ref{mass3-fig} for the region
except for the forbidden range, $2.11 < m_q < 2.68$, which has been found
and shown in Fig.~\ref{wq1fig} through the solutions of $w$.
In Fig.~\ref{mass5-fig}, the results for $q=10$ are shown. In this case,
we could find the mass gap at the transition point of
$m_q$ (shown by dashed line), which is
found through the solutions of $w$ (see Fig.~\ref{wq12fig}).

\vspace{.3cm}
The next problem is to compare the mass eigenvalues $m_9$ and $m_8$ obtained
with 
$2\tilde m_q$ as mentioned above
in order to see the stability of the meson states. 
In Figs.~\ref{mass2-fig}, \ref{mass4-fig}, \ref{mass6-fig}, 
$m_9$, $m_8$ and $E=2\tilde m_q$ are 
plotted with respect to $r_0=\sqrt{\lambda}/2$. In the case of $q=0$ 
(Fig.~\ref{mass2-fig}), $m_9$ and $m_8$ are larger than $2\tilde m_q$
at any point of ${\lambda}$. Then there is no stable meson state for $q=0$.
This is reasonable since the quark is not confined at $\lambda=0$ for $q=0$
and positive $\lambda$ works to destroy the confinement phase.
On the other hand, for $q>0$, the quark is confined at $\lambda=0$, then we
expect stable mesons for small $\lambda$. 
In fact, in the case of $q>0$, we could find 
the small 
$\lambda$ region where the meson masses are smaller than $2\tilde m_q$.
From Figs.~\ref{mass4-fig} and \ref{mass6-fig}, the stable regions
are found as $r_0=\sqrt{\lambda}/2 < 0.1\ (0.2)$ for $q=0.3$ ($q=10$).
The critical value (or the upper bound) of $\lambda$ increases with $q$
as expected.

In our present universe, the value of $q^{1/4}$ is expected as the order of
hadron mass since it provides the QCD string tension \cite{GY}. Meanwhile,
$\lambda$ is expected to be very small from the
recent observation of small acceleration even if it exists. Then
the mesons in the present world would not be affected by this small $\lambda$
and all the mesons are stable. On the other hand, there would be no hadrons
in the early universe which is inflationally expanding under a large $\lambda$.
And the quark and gluons would move almost freely in spite of their strong 
interactions which could combine them to make a bound state at $\lambda=0$.

As for the vector field, its
equation of motion is obtained from the 
action (\ref{D7-action}) as \cite{BGN}
\beq
\partial_a(\sqrt{-\det g_{cd}} \, F^{ab})-
\frac{4\rho(\rho^2+w^2)}{R^4}\varepsilon^{bjk} \partial_j A_k=0 \,,
\label{geom}
\eeq
where $\varepsilon^{ijk}$ is a tensor density (i.e., it takes values
$\pm 1$). 
The second term comes from the
Wess-Zumino part of the action, proportional to the pullback
of the RR five-form field strength, and is present only if $b$ is one of
the $S^3$ indices.  And $g_{ab}$ are metric in the Einstein frame.  

Because of the existence of the cosmological constant, Lorentz invariance 
is broken. So, the resulting equations are different for a time 
component, space components and the components with $S^3$ indices even if 
$A_\rho=0$ gauge is taken.  But the final results are expected to be not so 
different from the scalar meson spectra.  So, we abbreviate the analysis 
of vectors here.

\subsection{Baryon}

It has been shown  
that baryons correspond to D5-branes wrapped around the 
compact manifold $M_5$ \cite{Gross,Witten}. 
Here we assume it to be $S^5$. The brane action of 
such a D5 probe is 
\beq
S_{\rm D5}= -\tau_5 \int d^6\xi e^{-\Phi} \sqrt{\cal G} \ ,
\label{D5-action}
\eeq
where $(\xi_i)=(X^0,X^5 \sim X^9)$, $\tau_5$ represents 
the tension of D5 brane, and 
${\cal G}=-{\rm det}({\cal G}_{i,j})$ for the induced metric 
${\cal G}_{ij}= \partial_{\xi^i} X^M\partial_{\xi^j} X^N G_{MN}$. 
The mass of the wrapped D5-brane is then 
\bea
M_{\rm D5}(r)=\tau_5 e^{-\Phi} \sqrt{\cal G} =
\tau_5 \pi^3 R^4 r A(r) e^{\Phi/2} \; .
\eea
Before seeing the $\lambda$ dependence of this quantity, we consider the
case of $\lambda=0$, 
\beq
M_{\rm D5}(r)=\tau_5 \pi^3 R^4 r \sqrt{1+ {q \over r^4}} \; .
\eeq
This 
has a global minimum $M_{\rm D5}(r_{\rm min})=\tau_5 \pi^3 R^4 (4q)^{1/4}$,
which is regarded as the baryon mass,
at $r=r_{\rm min}=q^{1/4}$.
Thus, the baryon mass is also
induced by the $q$, i.e. by the gauge-field condensate. 
This is consistent with the fact that the QCD string tension is given by 
$q$ in the present model.

\begin{figure}[htbp]
\begin{center}
\voffset=15cm
  \includegraphics[width=9cm,height=7cm]{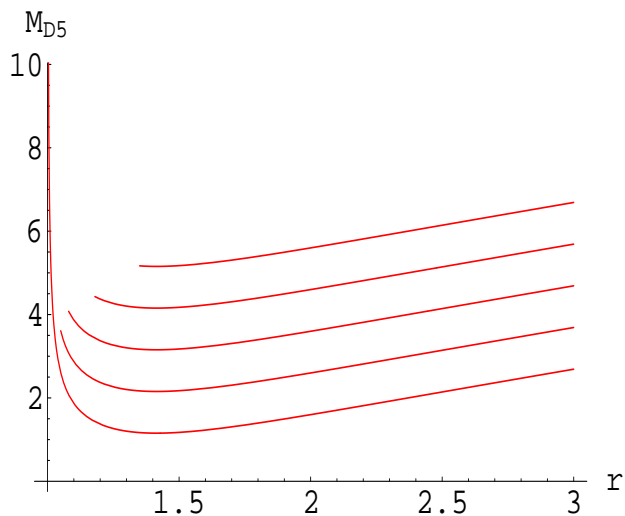} 
\caption{The D5-brane mass $M_{\rm D5}(r)$ as a function of $r$. 
Here we set $q=1$, $R=1$ and $\tau_5=1/\pi^3R^4$.  
The curves show the of $M_{\rm D5}|_{\lambda=4}$, 
$M_{\rm D5}|_{\lambda=4.41}+1$, $M_{\rm D5}|_{\lambda=4.67}+2$, 
$M_{\rm D5}|_{\lambda=5.57}+3$ and $M_{\rm D5}|_{\lambda=7.29}+4$
respectively, from bottom to up. 
\label{D5mass} }
\end{center}
\end{figure}

Fig.~\ref{D5mass} shows the $r$ dependence of $M_{\rm D5}(r)$ 
for five values of $\lambda$. 
For small side of $\lambda$, there exists a minimum 
at an appropriate point of $r$, and it disappears for large
enough value of $\lambda$. Thus, there is a transition
point of $\lambda$ of confinement for the baryon as seen in the
meson case. 

\section{Summary}

The Yang-Mills theory with light flavor quarks is investigated 
in the inflationally expanding 4d space-time or in the
time-dependent dS${}_4$ space-time.
The flavor quarks are introduced by embedding the D7 brane as a probe
in the background of the dual supergravity. The 10d background is 
deformed from AdS${}_5\times S^5$ by the dilaton and axion, and its 4d
boundary of AdS${}_5$ is set as the time-dependent dS${}_4$ space-time 
with a 4d
cosmological constant which is given as
an arbitrary constant parameter in our model.

In this model, the conformal invariance is broken even at the ultraviolet
boundary, then, in obtaining
the D7 embedding, the asymptotic form of the profile function $w(\rho)$
must include 
logarithmic terms coming from the loop-corrections.
Here we find the following form 
$$  w(\rho) \sim m_q+{c_0-4m_q r_0^2\log(\rho)\over \rho^2}, $$
at the lowest order of large $\rho$ limit,
and this implies the vev of the bilinear of quark fields, $\la\bar{\Psi}{\Psi}\ra$,
receives the loop correction proportional to the cosmological constant
$\lambda$ and the quark mass $m_q$ as
\beq
 -\la\bar{\Psi}{\Psi}\ra={c\over R^4}={c_0\over R^4}-m_q \lambda \log(\rho).
\eeq
This kind of correction would be expected in other quantities also, 
and we like to turn on this point in the future. 

In terms of the above asymptotic form, 
we obtain various solutions of $w(\rho)$ with different
values of $m_q$ and $c_0$. And we find
$c < 0$ for any solution of $m_q>0$ and $c=0$ for $m_q=0$. This implies that
the chiral symmetry is kept being unbroken for dS${}_4$. The solutions are 
separated to two groups by their infrared end point whether it is above
the horizon or just on the horizon. And when the
solution is switched from the one group to the other, three kinds of
behaviors are observed. For $q=0$, the transition of the solutions is smooth,
but there is a forbidden region of $m_q$ for the case of small $q$. For
large $q>0$, on the other hand, we find a phase transition, at an appropriate
value of $m_q$, and this is similar to the
one observed in the finite temperature gauge theory.

\vspace{.3cm}
In order to see the quark confinement, the Wilson-Polyakov loops are studied.
Our model for $q>0$ shows the quark confinement at $\lambda=0$ since
we find a linear rising potential with respect to 
the distance between quark and anti-quark, $L$, in this case. 
On the other hand, in dS${}_4$ or for $\lambda>0$, the
potential increases with $L$, but it
disappears for large $L$, $L>L_{max}$. 
And we find that the energy of quark and 
anti-quark system at $L=L_{max}$ is equal to the one of two parallel strings,
which connect horizon and the D7 brane. This means that the 
quark and anti-quark do not make the bound state any more for $L>L_{max}$
and they move freely and independently.
In this sense, we can say that
the gauge theory in dS${}_4$ is in the quark deconfinement phase.

We should notice that, in the present finite $\lambda$ case, 
the U-shaped string configuration is unique at a
given value of $L$.
This point is in sharp contrast to the case of finite temperature case, 
where there are two possible U-shaped
string configurations at the same values of $L$. However, the energy of 
one of them is always higher than the one of the two parallel string 
configuration. In this sense, the stable U-shaped string configuration is
unique in both cases, i.e. in the finite temperature and finite 
$\lambda$ cases.

\vspace{.3cm}
While the gauge theory of dS${}_4$ is in the deconfinement phase, we expect
that some meson states are stable for
small $\lambda$ due to the following reasons.
The value of $L_{max}$ proportional to $\lambda^{-1/2}$, so we will find
a U-shaped string with long $L$ 
at small $\lambda$.
Secondly, the energy of the parallel string configuration 
$2\tilde{m}_q$ 
increases with decreasing $\lambda$. Actually,
$\tilde{m}_{q}$ becomes infinite at $\lambda=0$ for finite $q$. Then at small
$\lambda$, $2\tilde{m}_q$ exceeds the meson mass considered.
In order to assure this point, 
the spectra of mesons are examined through the fluctuations
of D7 brane. Then, we can show
that any meson state for a definite quark
mass becomes stable when we take $\lambda$ to be small enough.

As for the baryon, which is identified as D5 brane wrapped on $S^5$,
its energy is obtained by the D5 brane action.
We could find that 
their mass is induced by the gauge condensate $q$ 
in our model.  For finite $\lambda$,
in the small side of $\lambda$, there exists a minimum of the D5 energy
at an appropriate point of $r$, and it disappears at large
value of $\lambda$. In this sense, we expect a transition
point of $\lambda$ of quark confinement for the baryon. The details of the
transition point will be given in the future work.
                                                                                
\vspace{.3cm}
\section*{Acknowledgments}

This work has been supported in part by the Grants-in-Aid for
Scientific Research (13135223)
of the Ministry of Education, Science, Sports, and Culture of Japan.

\vspace{1cm}
\section*{Appendix A}

We start form the type IIB supergravity with the following bosonic action,
\beq
 S={1\over 2\kappa^2}\int d^10x\sqrt{-g}\left(R-
{1\over 2}(\partial \Phi)^2+{1\over 2}e^{2\Phi}(\partial \chi)^2
-{1\over 4\cdot 5!}F_{(5)}^2
\right), \label{10d-action}
\eeq
where other fields are neglected since we need not them. By taking the
ansatz for $F_{(5)}$, 
$F_{\mu_1\cdots\mu_5}=-\sqrt{\Lambda}/2~\epsilon_{\mu_1\cdots\mu_5}$ 
\cite{KS2,LT}, and for the 10d metric as $M_5\times S^5$ or
$ds^2=g_{MN}dx^Mdx^N+g_{ij}dx^idx^j$, the equations of motion are given as
\bea
 && R_{MN}=-\Lambda g_{MN}+{1\over 2}\partial_M \Phi\partial_N \Phi
   -{1\over 2}e^{2\Phi}\partial_M \chi\partial_N \chi \, , \label{5metric} \\
  &{}&{1\over \sqrt{-g}}\partial_M\left(\sqrt{-g}g^{MN}\partial_N\Phi
\right)=-e^{2\Phi}\partial_M \chi\partial_N \chi g^{MN} \, , \label{phieq} \\
  &{}&{1\over \sqrt{-g}}\partial_M\left(\sqrt{-g}g^{MN}e^{2\Phi}\partial_N\chi
\right)=0 \, , \label{chi-eq} \\
  && ~~~~~~~~~~~~~~~~~~R_{ij}=\Lambda g_{ij} \, .
\eea
Using the ansatz, $\chi=-e^{-\Phi}+$ constant, the equation (\ref{5metric})
is written as 
$$R_{MN}=-\Lambda g_{MN}.$$
Then the metric of $M_5$ part is solved in the following form \cite{BGOY}
\beq
 ds_{(5)}^2={r^2 \over R^2}A^2\left(-dt^2+a(t)^2(dx^i)^2\right)+
\frac{R^2}{r^2} dr^2
\eeq
\beq
  A=1-({r_0\over r})^2, \quad a(t)=e^{2{r_0\over R^2} t}.
\eeq
Using this result, $\Phi$ in the Eqs.(\ref{phieq}) and (\ref{chi-eq}) are 
solved as given in the section 2. We should notice that the integration
constants are set appropriately in solving the equations.

Our model is based on type IIB supergravity, and we solved the equations
of motion with dilaton and axion. In the present case, the model is reduced
to 5d gauged supergravity. However, the solution given above breaks the 
supersymmetry since it is not a solution of the first order equations, which
are written in terms of the superpotential $W(\Phi, \chi)$, which is a constant
in the present case, of the 5d theory
\cite{FGPW}. This is consistent with the fact that there is no supersymmetric
theory in dS${}_4$. We notice that
the supersymmetry is also broken by our D7 brane embedding 
since the $\kappa$ symmetry of the D7 brane action is lost \cite{GY}.

\vspace{.5cm}
\section*{Appendix B}
Here we show the constraint given in Eq.(\ref{constraint}) in the Sec.2. Suppose
some solution $w(\rho)$ of the group (b), then its Infrared end point 
sits on the horizon at $\rho=\rho_0$. Then we expand $w(\rho)$ near the 
end point as
\beq
 w(\rho_0+\epsilon)=w(\rho_0)+w'(\rho_0) \epsilon+\cdots \, .
\eeq
By substituting this asymptotic form into the equation of $w$, Eq.(\ref{qeq}),
we obtain
\beq
 w'(\rho_0)={\rho_0+3\sqrt{r_0^2-9w_0^2}\over r_0^2-10w_0^2}w_0 \, ,
\eeq
where $w_0=w(\rho_0)$ and we used $\rho_0^2=r_0^2-w_0^2$. Since the physical 
solutions should be monotonically increasing with $\rho$ in the present case, 
then we obtain $w_0<r_0/\sqrt{10}$. This constraint does not depend on the 
value of $q$, while this does not appear for $q=0$. In other words, this 
behaviour is given by the non-trivial dilaton or the gauge condensate in the
corresponding 4d gauge theory.


\newpage
\end{document}